\documentstyle[prl,aps]{revtex} 
\begin{document}
\draft

\def\pdov#1#2{{\partial#1\over \partial#2}}

\twocolumn[
\hsize\textwidth\columnwidth\hsize\csname@twocolumnfalse\endcsname
\title{\bf THE ISSUE OF CHOOSING NOTHING: WHAT DETERMINES 
THE LOW ENERGY VACUUM STATE OF NATURE?}
\author{T. Padmanabhan$^*$ and T. Roy Choudhury$^{\dagger}$}
\address{IUCAA, Post Bag 4, Ganeshkhind, Pune 411 007, India.\\}
\date{\today}
\maketitle 
\begin{abstract}
Starting from an (unknown) quantum gravitational model, one can invoke 
a sequence
of approximations to progressively arrive at quantum field theory (QFT) 
in curved spacetime, QFT in flat spacetime, 
nonrelativistic quantum mechanics and newtonian mechanics. 
The more exact theory can put restrictions on the range of possibilities
allowed for the approximate theory which are not derivable from  the latter --
an example being the symmetry restrictions on the wave function for a pair
of electrons. We argue that the choice of vacuum state at low energies could
be such a `relic' arising from combining the principles of quantum theory 
and general
relativity, and demonstrate this result in a simple toy model. 
Our analysis suggests that the wave function of the universe, 
when it describes the large volume limit of the universe, 
dynamically selects a vacuum state for matter fields --- which in 
turn defines the concept of particle in the low energy limit.
The result also has the 
potential for providing a concrete quantum mechanical version of 
Mach's principle.
\end{abstract}
% for PACS codes, see http://publish.aps.org/PACS/pacs99.html
]
\narrowtext

It is a well known fact that our knowledge about the fundamental 
laws of physics becomes
increasingly uncertain, as we proceed to higher energies. A fortunate 
feature about 
Nature seems to be  that the high energy behaviour of physical systems 
do not influence the low
energy predictions based on effective hamiltonians, allowing a 
steady progression of our 
understanding through a sequence of effective theories. This behaviour,
which can be formalised in terms of renormalisation group analysis, 
forms the cornerstone of quantum field theoretical description 
of Nature at low energies. 

There are, however, some key features of
high energy phenomena which leave traces at low energies as regards 
the restrictions
on allowed quantum states. For example, the Pauli
exclusion principle plays a vital role in low energy atomic physics 
but cannot be derived
or even related to any feature of low energy hamiltonian -- its origin 
lies in relativistic
field theory. Within the framework of low energy theory, there is no 
symmetry restrictions
on the quantum state for, say, a system of two electrons. But if the 
electrons are treated
as excited states of an underlying fermionic field, it is possible 
to prove that
only a subset of quantum states --- which are antisymmetric --- are 
allowed. The importance of such `relic principles' lies 
in providing a glimpse of the unknown
territory. 

This prompts us to ask: Is there any such effect 
or selection principle which 
arises from the high energy theory that combines 
general relativity and quantum theory but leaves a trace
at low energies? We argue in this Letter that the choice 
of the vacuum state of low energy could itself
be a relic of an intrinsically quantum gravitational principle.

We begin by highlighting certain issues which arise in the 
definition of vacuum state
(and one-particle state) in the low energy theory when 
viewed along conventional lines.
In nonrelativistic, newtonian, mechanics the concept of a 
free particle moving with
uniform velocity in an inertial frame, also presupposes the existence of a 
unique {\it time} coordinate. The
equation $d^2{\bf x}/dt^2=0$ is invariant only under the 
linear transformation $t\to
\alpha t +\beta$. This feature continues in 
nonrelativistic point quantum mechanics.
At the next level, combining the principles of special 
relativity with quantum theory
extends the accepted range of time coordinate to 
that of any inertial observer given by 
standard Lorentz transformations. This
class allows one to define a set of positive frequency modes, 
creation and annihilation
operators and the inertial vacuum state $|0\rangle_I$. 
Conventional QFT works at this level and
the choice of vacuum state is now unique because 
the class of inertial frames form a
privileged set in special relativity.

This description, however, is unacceptable both conceptually 
and technically at a fundamental level.
Conceptually, there is no reason to attribute a 
special status to any one class of observers
--- a point forcefully emphasised by Einstein while 
motivating general relativity~\cite{eins}. Technically, 
any quantum field generates a gravitational field around it 
thereby curving the spacetime; a
description of QFT in flat spacetime can only be an approximation 
and no fundamental feature of
the theory (like the choice of vacuum state or definition of 
particles) should depend on the
approximate model.  Doing QFT in curved spacetime is 
{\it not} a matter of choice but {\it is
mandatory, because all fields curve the spacetime}. 
This changes the symmetry group to that of arbitrary coordinate 
transformations and extends the allowed choices of
time coordinate. The uniqueness of vacuum state 
(or the definition of particle) is lost
when we work in the limit of QFT in a general curved spacetime~\cite{bide}.

Even this situation, however, is unacceptable because we 
cannot really address the issue
of ground state when gravity is treated as  $c$-number field and 
the matter field is quantised.
One should really consider the full Hilbert space of the quantum 
theory of matter field coupled
to gravity, (we are essentially interested in the self gravity of matter; 
but the formalism
remains the same) and try to define a ground state for this theory. 
The structure of such a
theory is at present unknown but the indications 
are that the ground state wave functional
will have an extremely complex structure -- 
since it incorporates the arbitrarily high energy
virtual modes of not only the matter fields but also gravity 
(which is nonlinearly coupled to itself). 
If the fundamental description is in terms of strings or 
spin networks, it is not
even clear how to address the question of ground state in the 
language of conventional QFT~\cite{pol}.
But the above arguments clearly suggest that the choice of 
ground state for any quantum
field theory is linked with very high energy phenomena.

It is possible make some progress if we observe that the issue of ground state
has implications for cosmology. Note that an arbitrary
choice of the time coordinate, even in newtonian mechanics, 
will make a free particle
move in an accelerated trajectory.  The result persists 
even in QFT done in a noninertial frame. It is easy
to show that when the nonrelativistic wave function 
corresponding to the particles defined
in such a coordinate system is taken, it will have 
psuedopotential term giving raise to an 
accelerated motion~\cite{pad94}. 
For example, let us consider a particle, of mass $m$,  
described by a quantum field $\hat{\phi}$ in flat spacetime.
Let $|0\rangle_I$ and $|1_{\bf k}\rangle_I$ denote the vacuum and one-particle 
states for the field in the inertial frame. The nonrelativistic 
limit ($c \to \infty$) of the QFT can be obtained by identifying the quantity 
$_I\langle0|\phi|1_{\bf k}\rangle_I$ with the 
Schr\"odinger wave function $\psi$. It can then be shown that $\psi$ 
obeys the free particle Schr\"odinger equation. However, if we work in 
a noninertial frame, say, the Rindler frame, given by the metric 
\begin{equation}
ds^2=\left(1+\frac{gx}{c^2}\right)^2 c^2 d\tau^2 -dx^2-dy^2-dz^2,
\end{equation}
then the particle states 
$|0\rangle_R$ and $|1_{\bf k}\rangle_R$, 
defined using the Rindler time coordinate $\tau$,
will not be the same as those defined using the inertial time $t$. 
In the nonrelativistic limit, the function 
$_R\langle0|\phi|1_{\bf k}\rangle_R$ has to be identified with 
the wave function $\psi$, which will satisfy the 
Schr\"odinger equation for a uniformly accelerated particle~\cite{pad94}
\begin{equation}
i \hbar \pdov{\psi}{\tau}=-\frac{\hbar ^2}{2 m} \nabla^2 \psi+mgx\psi.
\end{equation}
In the newtonian limit $(\hbar \to 0$), this equation will 
describe a particle moving under a pseudopotential $gx$. Hence, if we 
choose our time coordinate as the Rindler time $\tau$, 
the particles (defined using $\tau$), in the nonrelativistic limit, 
will experience a pseudo force.
To realise that there is a pseudo force, one has to
introduce some kind of ``fixed frame of stars'' as is usually 
done in discussions of Mach's
principle which --- in a more sophisticated form --- 
will be connected to the boundary
conditions on the wave function describing the state 
of the universe~\cite{isen}.

To see this explicitly, we shall assume that the conventional
approach to quantum cosmology 
(say, based on Wheeler-DeWitt equation) does interface between
the fully quantum gravitational description of spacetime 
(say, in terms of spin networks or strings) 
and conventional QFT. Then the question of ground 
state translates to finding an
acceptable solution to Wheeler-DeWitt equation, $\Psi(g,\phi)$, 
depending on both gravitational
and matter variables denoted symbolically as $(g,\phi)$. 
Conventionally, Wheeler-DeWitt equation
is solved in a FRW minisuperspace, in which a choice of 
time coordinate is already made --- which
defeats our purpose. To test our conjecture that 
quantum cosmological solution can effect a choice
of vacuum state we need to find a sufficiently general 
but yet tractable approximation to Wheeler-DeWitt equation. 
Fortunately, this can be done along the following lines:

The least amount of dynamical structure needed to illustrate 
our idea is provided by the Bianchi
Type~I minisuperspace, with the metric
\begin{equation}
\begin{array}{ccl}
ds^2&=&e^{-6\Omega}dt^2\nonumber\\
    &-&e^{-2\Omega}[e^{2(\beta_+ + \sqrt{3} \beta_-)}dx^2\nonumber\\
    &+&e^{2(\beta_+ - \sqrt{3} \beta_-)}dy^2\nonumber\\
    &+&e^{-4\beta_+}dz^2].
\end{array} 
\end{equation}
The classical dynamics of the Bianchi Type~I empty universe is described by the
Kasner solutions~\cite{ll}
\begin{equation}
\Omega \propto t,~\beta_+=C_+\Omega,~\beta_-=C_-\Omega
\label{eq:kasner}
\end{equation}
with the constraint
\begin{equation}
C_+^2+C_-^2=1.
\label{eq:constraint}
\end{equation}

There is an equivalent description of the empty Bianchi Type~I universe 
for which the metric has the form~\cite{ll}
\begin{equation}
ds^2=dT^2-T^{2 p_1}dx^2-T^{2 p_2}dy^2-T^{2 p_3}dz^2,
\end{equation}
with the constraints 
\begin{equation}
p_1+p_2+p_3=p_1^2+p_2^2+p_3^2=1.
\end{equation}
The time coordinates are related by 
\begin{equation}
dT=e^{t}dt,
\end{equation}
where we have chosen the scale of $t$ in such a way that $\Omega = -t/3$, 
while the 
quantities $p_1,p_2,p_3$ are related to $C_+,C_-$ through the relations 
\begin{equation}
\begin{array}{ccl}
p_1 &=& \frac{1}{3} (1-C_+-\sqrt{3}C_-);\\
p_2 &=& \frac{1}{3} (1-C_++\sqrt{3}C_-);\\
p_3 &=& \frac{1}{3} (1+2C_+).
\end{array}
\end{equation}
There is also a special solution where $p_1=p_2=p_3=0$, which describes 
the Minkowski metric.

The solutions given by equations (\ref{eq:kasner}) and (\ref{eq:constraint}), 
in general, 
represent curved spacetimes, even though they are source free.
However, there exist a subset of flat spacetime solutions among them 
which we shall concentrate on
for our illustration. The key point to note is that there are {\it two 
distinct classes of flat spacetimes.} 
The first class is given by the three solutions 
\begin{equation}
{\rm Class~I:~ } \left \{ \begin{array}{l}
(i)~C_+=-\frac{1}{2}, C_-=-\frac{\sqrt{3}}{2};\\
(ii)~C_+=-\frac{1}{2}, C_-=\frac{\sqrt{3}}{2};\\
(iii)~C_+=1, C_-=0.
\end{array} \right.
\label{eq:class1}
\end{equation}
which are the flat Milne universes~\cite{milne} and differ 
only in the choice of spatial direction
along which expansion takes place. The solutions in terms of $p_1,p_2,p_3$, 
are given by 
\begin{equation}
\begin{array}{c}
(i)~p_1=1,p_2=0,p_3=0;\\
(ii)~p_1=0,p_2=1,p_3=0;\\
(iii)~p_1=0,p_2=0,p_3=1.
\end{array}
\end{equation}
The second class corresponds to the choice  
\begin{equation}
{\rm Class~II:~} \Omega=\beta_+=\beta_-=0, 
\label{eq:class2}
\end{equation}
which describes the flat Minkowski universe.
The classical dynamics of this system [given by equations (\ref{eq:kasner}) 
and (\ref{eq:constraint})] 
can be described geometrically in the
minisuperspace for this metric, which is the 
$(\Omega, \beta_+,\beta_-)$-space. The 
Kasner solutions are straight lines which lie
on a ``light cone'' like structure. The three Milne flat solutions 
are three straight
lines on this light cone separated by $120^{\circ}$. 
{\it The Minkowski flat universe, 
on the other hand, lies at the origin.} 
Since we have three variables $(\Omega, \beta_+,\beta_-)$ here, 
we can choose $\Omega$
as our time coordinate. The other two variables can then act as our dynamical
variables.

The key point to note is the following:  The two distinct class of 
solutions described above in equations (\ref{eq:class1}) 
and (\ref{eq:class2}) both correspond
to flat spacetimes but differ in the choice of time coordinate. 
In quantum theory, they
would correspond to different choices of vacuum states~\cite{pad90}. 
By constructing the necessary  
quantum description we can investigate how the wave function 
of the universe behaves vis-a-vis the choice of ground state.

To describe the quantum dynamics, we start with the Wheeler-Dewitt equation,
which takes a particularly simple form for this metric~\cite{dewitt}:
\begin{equation}
\pdov{^2 \psi}{ \Omega^2}-\pdov{ ^2 \psi}{\beta_+^2}
             -\pdov{ ^2 \psi}{ \beta_-^2}=0.
\label{eq:wd}
\end{equation}
As usual, the Wheeler-DeWitt equation is ``timeless'', 
but we can take $\Omega$ as our
a fiducial evolutionary parameter. The relevant 
solution for this equation, which gives a positive definite 
probability density, and also describes an expanding universe, 
can be written in the form
\begin{equation}
\psi(\Omega,{\bf r})=\int K_+({\bf p})e^{i({\bf p} \cdot {\bf r} 
                               -|{\bf p}| \Omega)} d^2p,
\label{eq:soln}
\end{equation}
where we have introduced the two-dimensional vectors
\begin{equation}
{\bf r}=(\beta_+,\beta_-),~{\bf p}=(p_+,p_-).
\end{equation}
Since equation (\ref{eq:wd}) is second order in time, the probability 
density $|\psi(\Omega,{\bf r})|^2$ is, in general, not positive definite. 
This problem is well known for Wheeler-Dewitt equation, and is discussed 
by several authors (see~\cite{dewitt}). In this work, we need not worry 
about this issue because the probability for the solution (\ref{eq:soln}) 
is always positive.  
Let us assume that $|\psi(\Omega,{\bf r})|^2$ is peaked 
around the Minkowski vacuum state initially. This means we can choose
\begin{equation}
\psi(0,{\bf r})=f({\bf r}) e^{i {\bf q} \cdot {\bf r}}
\end{equation}
where $f({\bf r})$ is peaked around ${\bf r}=0$. As $\Omega$ increases, the 
wave function will spread,
and the peak will move. At some sufficiently large $\Omega$, 
the probability distribution, $|\psi(\Omega,{\bf r})|^2$ will 
be peaked around one of the classical trajectories, 
propagated along the characteristics of equation (\ref{eq:wd}).
The direction of ${\bf q}$ will decide the specific 
classical solution around which the probability
will be peaked. At some given $\Omega$, the probability will be
peaked around the point 
\begin{equation}
{\bf r}=\frac{\bf q}{|{\bf q}|} \Omega.
\end{equation}
In the case where ${\bf q}=0$, the probability, 
instead of peaking around any specific Kasner solution, will 
be peaked around the whole class of Kasner solutions, i.e., 
for a given 
$\Omega$, it will be peaked around the circle 
$r^2 \equiv \beta_+^2+\beta_-^2=\Omega^2$. Hence, it can 
be concluded that the quantum solutions at large $\Omega$ will always  
deviate from the Minkowski universe. 

In order to describe the ground state let us 
concentrate on the flat spacetime solutions. Hence we
choose our initial conditions (i.e., the direction of ${\bf q}$) 
in such a way that when $\Omega > 0$, the
probability is peaked around any one  of the Milne flat solutions. 
We thus see that even if we localise the wave function around 
the Minkowski universe
initially, it will be localised around one of the Milne 
solutions at later stages.  
The quantum  wave function described here tend to pick up the
Milne universe at late times -- hence the Milne time coordinate 
is preferred over the Minkowski one.
We can now introduce matter fields with spatial 
degrees of freedom, take the limit of
semiclassical gravity~\cite{qftcst} to obtain QFT in curved spacetime and 
proceed to low energy limit. 
The Wheeler-DeWitt equation for the full system is nothing but 
the hamiltonian constraint equation
\begin{equation}
(H+H_{\rm matter})\Psi(g,\phi)=0.
\end{equation}
We write $\Psi(g,\phi)$ as 
\begin{equation}
\Psi(g,\phi)=R(g,\phi) \exp[\frac{i}{\hbar} S(g,\phi)],
\end{equation}
substitute it in the hamiltonian constraint equation and   
take the limit $\hbar \to 0$. Now, if we 
identify the derivative of the action $S$
with respect to the gravitational variables (written symbolically as
$\delta S/\delta g$) with the canonical momentum, 
we will get the usual dynamical equations 
for free gravity, and, in addition, 
$R$ will satisfy the Schr\"odinger equation
\begin{equation}
i \hbar \frac{\delta S}{\delta g}\frac{\delta R}{\delta g}
= H_{\rm matter} R.
\end{equation}
This equation represents the QFT in curved spacetime. We can identify the 
time coordinate $\tau$ through the derivative of the action $S$
\begin{equation}
\frac{d}{d\tau}=\frac{\delta S}{\delta g}\frac{\delta}{\delta g}.
\end{equation}
It is
obvious that, in our case, the vacuum state and particles will be defined 
with respect to Milne time coordinate.

This simple -- but adequate -- example illustrate the 
idea we suggested earlier:
It is possible that the wave function of the universe, 
when it describes the large
volume limit of the universe (in our case large $\Omega$ limit), 
dynamically selects a vacuum state for matter fields --- which in 
turn defines the concept of particle in the low energy limit. 

This particular example leads to a vacuum state corresponding
to anisotropic expansion, which is --- of course --- not a
proper description of low energy physics. But it should clear that
this particular feature is an artifact of the minisuperspace we
have chosen.  
We stress the fact that the general arguments given in this Letter are 
of far greater
validity than the simplified example given above 
(which could be objected to at several levels).  
To emphasise the broader picture we conclude with
a summary of the key ingredients of
our analysis:
(1) All quantum fields curve the spacetime around them due to 
self gravity. Hence, QFT in
a flat spacetime is an approximation. 
(2) The choice of Minkowski (inertial) time coordinate
ceases to exist in  a curved spacetime. Thus the true ground 
state of any field is nontrivial   
when we realise that it generates a gravitational field around it. 
We need to ask whether
the full quantum theory of gravity plus matter fields puts 
restrictions on the choice of
possible ground states. The analogy will be with the allowed quantum states
for a pair of electrons in atomic physics. Nothing in low energy theory
prevents the existence of a state with arbitrary symmetry; but when we treat
electrons as low energy excitations of an underlying fermionic 
field theory, we find that
only antisymmetric states are allowed. 
(3) Since the ground state wave function of universe  also figures
prominently in the description of quantum cosmological models 
of the universe, there arises the 
interesting possibility of connecting the definition of free 
particle at low energies with
the geometry of the universe, thereby providing a quantum 
version of Mach's principle.
(4) A toy model  demonstrates that it is possible to 
have a dynamical system which has these features.

\bigskip
TRC acknowledges the financial support from 
the University Grants Commission, India.


\begin{thebibliography}{20}
\bibitem{milne}
We clarify the usage of the term Milne metric: 
In literature, the term has been used to 
describe two different metrics. Usually it is used to describe 
the $k=-1, a(t) \propto t$ FRW metric. However, in this Letter, the 
Milne metric has been used to describe the flat anisotropic 
universe which expands with $a(t) \propto t$ 
along only one particular spatial direction.
\end{thebibliography}
\end{document}